\documentclass[]{article}
\usepackage{amsmath}
\usepackage{amsfonts}
\usepackage{amssymb}
\usepackage{graphicx}
\usepackage{amsfonts}
\usepackage[english]{babel}
\usepackage[utf8]{inputenc}
\usepackage{times}
\usepackage[T1]{fontenc}

\begin{document}

\title{Convergent $\tilde{Y}$-Map for a new covariant Loop Quantum Gravity formulation}

\author{Leonid Perlov\\
Department of Physics, University of Massachusetts,  Boston\\
leonid.perlov@umb.edu
}
\date{ October 6, 2015}

\maketitle

\begin{abstract}
The most important part of the new spin-foam loop quantum gravity formulation $\cite{RovelliBook2}$, $\cite{Rovelli2010}$ is the map  $Y$: $H^{SU(2)} \rightarrow H^{SL(2,C)}$. It was only recently shown that the Y-Map is convergent $\cite{Perlov2015-2}$ in spite of the fact that the classical Peter-Weyl theorem is not applicable to it, as Lorentz group is not compact. In this paper we provide an alternative map $\tilde{Y}$. The $\tilde{Y}$  map has an advantage of preserving the Lorentz covariance, which gets broken in the case of Y-Map. The image of a new map $\tilde{Y}$ contains the weighted infinite sum of $SL(2,C)$ matrix coefficients. The sum is convergent and its limit is the square integrable functions of $SL(2,C)$  with the  measure  $L^2(g, e^{-|Y|^2/\hbar}\eta(g) du \,dY )$ according to the Holomorphic Huebschmann-Peter-Weyl theorem \cite{Huebschmann}, which is applicable to the rational representations of the non-unitary groups, particularly non-unitary finite Lorenz representations. Since in LQG the unitary evolution is not mandatory as it does not follow from the Wheeler-DeWitt dynamics equation, the choice of the non-unitary representation is valid. As it was stated in the original LQG formulation $\cite{RovelliBook}$ in the section 10.1.3  "there is no sense in which conventional unitarity is necessary in the theory".\\ 
\end{abstract}

\section{Introduction}

A new covariant loop quantum gravity formulation $\cite{RovelliBook2}$, $\cite{Rovelli2010}$ is based on the map $Y$ which is a map $H^{SU(2)} \rightarrow H^{SL(2,C)}$. $Y$ maps $H^{SU(2)}$ states $|j, q\rangle$ to $H^{SL(2,C)}$ states $|j, q> \rightarrow |j, j\gamma, j, q\rangle,  j\in Z^+, \gamma - \mbox{Barbero-Immirzi}$. it maps $SU(2)$ matrix coefficients $D^j_{qq'}(u)$  to $SL(2,C)$ unitary principal series matrix coefficients $D^{(j, j\gamma)}_{jqjq'}(g)$. By the classical Peter-Weyl theorem any $SU(2)$ function can be decomposed into the infinite sum of Wigner matrix coefficients $D^j_{qq'}(u)$. The $Y$ then maps the functions of SU(2) to the functions of SL(2,C) in the following manner:
\begin{equation} 
Y: \phi(u) = \sum\limits_{j=\frac{|m|}{2}} \sum\limits_{q,q' = -j}^{j} C_{jqq'} D^j_{qq'}(u) \rightarrow \psi(g) = \sum\limits_{j=\frac{|m|}{2}} \sum\limits_{q,q' = -j}^{j} C_{jqq'} D^{j, j\gamma}_{jqjq'}(g)
\end{equation}
where, $C_{jqq'}$ - are the $SU(2)$ Fourier transform of $\phi(u)$. \\
The authors of the new LQG formulation $\cite{RovelliBook2}$, $\cite{Rovelli2010}$ do not discuss the convergence or the meaning of this sum, even though it is known that generally it is impossible to expand the function on the non-compact Lorentz group into the sum of its matrix coefficients of the unitary representation due to the lack of the Peter-Weyl theorem for the unitary representations of the non-compact groups.  However, as it was recently proved in $\cite{Perlov2015-2}$ the sums are really convergent in spite of the fact that the Peter-Weyl theorem is not applicable. The limit of the sums however are not square integrable functions and therefore it is impossible to directly introduce the inner product on them. Thus the Y-Map must use the 3-dimensional projected space and define the inner product on the 4 dimensional space to be equal to the inner product on the 3 dimensional hypersurface. This procedure necessary breaks the Lorentz covariance. \\
In this paper we define an alternative convergent map, which we call $\tilde{Y}$. We prove that $\tilde{Y}$ converges and the limit is a square integrable function of SL(2,C) with the measure $(e^{-|Y|^2/\hbar}\eta(g) du \,dY ), g \in SL(2,C), u \in SU(2), Y \in su(2)$ . Therefore we can introduce the inner product directly without breaking the Lorentz covariance. We use the Holomorphic Huebschmann-Peter-Weyl theorem  \cite{Huebschmann} (J. Huebschmann 2008). which is applicable to the rational representations of the non-compact groups (non-unitary finite representation of the Lorentz group). Thus we obtain the Lorentz covariance at the price of unitarity. However unitarity is not mandatory in LQG as it does not follow from the Wheeler-DeWitt dynamics equation in the same manner as it follows in conventional QM from the Schrödinger equation.\\
As it was stated in the original LQG formulation $\cite{RovelliBook}$  in the section 10.1.3  "In conventional QM and QFT, unitarity is a consequence of the time translation symmetry of the dynamics. In GR there isn't, in general, an analogous notion of time translation symmetry. Therefore there is no sense in which conventional unitarity is necessary in the theory. One often hears that without unitarity a theory is inconsistent. This is a misunderstanding that follows from the erroneous assumption that all physical theories are symmetric under time translations."\\
The paper is organized as follows. In section \ref{sec:HolomorphicPeterWeyl} we state and discuss the Holomorphic Huebschmann-Peter-Weyl theorem.  In the subsequent section \ref{sec:ANewYMap} we introduce the convergent $\tilde{Y}$-Map and prove its convergence and square integrability.  The discussion section \ref{sec:Discussion} concludes the paper.

\section{Holomorphic Huebschmann-Peter-Weyl Theorem}
\label{sec:HolomorphicPeterWeyl}

The holomorphic Huebschmann-Peter-Weyl theorem \cite{Huebschmann} establishes the isomorphism between the Hilbert space spanned by the compact group K matrix coefficients and the Hilbert space spanned by the rational representation matrix coefficients of that group complexification $K^{\mathbb{C}}$:
\begin{equation}
\label{Holomorphicmap}
\phi^{\mathbb{C}}(g)  \rightarrow  {(\hbar\pi)}^{{dim(K)}/4}e^{\hbar{|\lambda + \rho|}^2/2}\phi(g)
\end{equation}
the inner products of these two Hilbert spaces are related by the following:
\begin{equation}
\label{eq:innerprodgeneral}
\int\limits_{K^{\mathbb{C}}}\bar\phi^{\mathbb{C}}(g)\phi^{\mathbb{C}}(g) e^{-|Y|^2/\hbar} \eta(g) du \,dY = {(\hbar\pi)}^{{dim(K)}/2}e^{\hbar{|\lambda + \rho|}^2} \int\limits_K \bar\phi(u)\phi(u)\, du
\end{equation}
, where $g \in K^{\mathbb{C}}$ and we use polar decomposition $g = ue^{iY}, u \in K, Y \in t, $ algebra of K,\; $\lambda$ - is the highest weight of $K$, while $\rho$ is the Weyl vector of $K$, i.e. the half sum of the positive roots, the density of the measure on the left hand side is
\begin{equation}
\label{eq:measure}
\eta(u, Y) = \left(det\left(\frac{sin(ad(Y))}{ad(Y)}\right)\right)^{\frac{1}{2}},  u \in K,  Y \in t
\end{equation}
It means that we can calculate the inner product in the non-compact group rational representation Hilbert space by calculating the inner product of its isomorphic projection to the Hilbert space of its maximum compact subgroup representation. And, what is important, it does not depend on the selection on the maximum compact subgroup, which means being $K^{\mathbb{C}}$ covariant. \\[2ex]
Since the map is provided by constant multiplication, all the orthonormal properties of the matrix coefficients in the compact group case propagate to the Hilbert space of its non-compact complexification.\\[2ex] 
Let $K$ be a compact group, $K^{\mathbb{C}}$ - its complexification, $t$ and $t^{\mathbb{C}}$ its algebras respectively. Let $g \in K^{\mathbb{C}}, u \in K, Y \in t$, $\eta(u, Y)$ as in (\ref{eq:measure}).
We denote as in  \cite{Huebschmann} ${\hat{K}}^{\mathbb{C}}$ to be the set of isomorphism classes of irreducible rational representations of $K^{\mathbb{C}}$. ${\hat{K}}^{\mathbb{C}}$  is identified with the space of highest weights corresponding to the  dominant Weyl chamber. For the highest weight $\lambda ,  \; T_{\lambda}$ is a rational representation in a class of $\lambda$: $K^{\mathbb{C}} \rightarrow \mbox{End}(V_{\lambda}), V_{\lambda}$ is a representation vector space. For $\psi \in V^{\star}_{\lambda}$ and $w \in V_{\lambda}$ the function $\varPhi_{\psi w}(q) = \psi(q w)$, $q \in K^{\mathbb{C}}$ is  a representative function on $K^{\mathbb{C}}$ and it provides a morphism:
$V_{\lambda}^{\star} \otimes V_{\lambda} \rightarrow \mathbb{C}[K^{\mathbb{C}}]$. We denote this morphism following  \cite{Huebschmann} as $V_{\lambda}^{\star} \odot V_{\lambda}$\\[2ex]
\textbf{Theorem} [Holomorphic Peter-Weyl J. Huebschmann 2008] \\[2ex]
The Hilbert space $HL^2(K^{\mathbb{C}},  e^{-|Y|^2/\hbar} \eta(g) du \,dY)$  contains the vector space $\mathbb{C}[K^{\mathbb{C}}]$ of representative functions (matrix coefficients) on $\mathbb{K^{\mathbb{C}}}$ as a dense subspace, and as a unitary  $(K \times K)$-representation, 
$HL^2(K^{\mathbb{C}},  e^{-|Y|^2/\hbar} \eta(g) du \,dY)$ decomposes as the direct sum into $K \times K$-isotypical summands:
\begin{equation}
HL^2(K^{\mathbb{C}},  e^{-|Y|^2/\hbar} \eta(g) du \,dY) = {\hat{\oplus}}_{\lambda \in {\hat{K}}^{\mathbb{C}}} V_{\lambda}^{\star} \odot V_{\lambda}  
\end{equation}\\[2ex]
\textbf{Theorem} [ J. Huebschmann 2008 \cite{Huebschmann} Theorem 5.3] \\[2ex]
The association: $ \phi^{\mathbb{C}}(g)  \rightarrow  {(\hbar\pi)}^{{dim(K)}/4}e^{\hbar{|\lambda + \rho|}^2/2}\phi(g)$ as $\lambda$ ranges over the highest weights induces a unitary isomorphism of unitary $(K \times K)$ representations. 
\begin{equation}
 HL^2(K^{\mathbb{C}},  e^{-|Y|^2/\hbar} \eta(g) du \,dY) \rightarrow L^2(K, dx)
\end{equation}
, where $\phi_{\lambda} \in V_{\lambda}^{\star} \odot V_{\lambda}, \lambda$ - is the highest weight of $K$, while $\rho$ is the Weyl vector of $K$, i.e the half sum of the positive roots. $dx$ is a Haar measure.\\[2ex]

For the details and the Theorem proofs see \cite{Huebschmann}. 
In this paper we apply these theorems to the case $K = SU(2), K^{\mathbb{C}}=SL(2,C)$ to derive a convergent $\tilde{Y}$ map.

\section{ $\tilde{Y}$-Map}
\label{sec:ANewYMap}

The Holomorphic Huebschmann-Peter-Weyl theorem establishes the isomorphism between the Hilbert space spanned by the compact group $K$ matrix coefficients and the Hilbert space spanned by the matrix coefficients of that group complexification $K^{\mathbb{C}}$
In this chapter we will use the Holomorphic Peter-Weyl theorem stated above in order to introduce $\tilde{Y}$ map. In our case $K$ is $SU(2)$, $K^{\mathbb{C}}$ is $SL(2,C)$. Let us derive for our case $\phi, \; \phi^{\mathbb{C}},  \; \lambda$, and $\rho$ and substitute them into (\ref{Holomorphicmap}).\\
\begin{equation}
\label{Holomorphicmap1}
\phi^{\mathbb{C}}(g)  \rightarrow  {(\hbar\pi)}^{{dim(K)}/4}e^{\hbar{|\lambda + \rho|}^2/2}\phi(g)
\end{equation}
The corresponding matrix coefficients $\phi$ and $\phi^{\mathbb{C}}$ are as follows:
\begin{equation}
\label{eq:representativefunction}
\phi(u) = D^j_{qq'}(u) , \qquad  \phi^{\mathbb{C}}(g) =  D^{(j_-, j_+)}_{ q q'}(g) =  D^{(j_-, 0)}_{ q q'}(g) \otimes  D^{(0, j_+)}_{ q q'}(\bar{g})
\end{equation}
where $j_+ - j_- = j$\\[2ex]

For $SU(2) \; \dim(K) = 3$, the highest weight $\lambda_j$ of the finite dimensional representation is $(\dim(V) - 1) \frac{\alpha(H)}{2}$, which is $2j \cdot \frac{\alpha(H)}{2}$, where $\alpha(H)$ is the only $SU(2)$ positive root $\alpha(H) = 2h$, $H = \mbox{diag}(ih, -ih)$. The Weyl vector $\rho = \frac{\alpha(H)}{2}$. The Killing form gives the value of ${|\lambda_j + \rho|}^2 =\frac{{(2j + 1)}^2}{8}$  By substituting these values into ($\ref{Holomorphicmap}$) we find the matrix coefficients map:

\begin{equation}
D^j_{qq'}(u)  \rightarrow \frac{1}{A_j} D^{(j_-, j_+)}_{qq'}(g), \;\;  j_+ - j_- =j
\end{equation} \\
, where 
\begin{equation}
\label{holomorphicoeff}
A_j = {(\hbar\pi)}^{3/4}e^{\frac{\hbar{(2j + 1)}^2}{8}}
\end{equation}
By the Holomorphic Peter-Weyl theorem $\frac{1}{A_j} D^{(j_-, j_+)}_{qq'}(g)$ are dense \\
in $L^2(g, e^{-|Y|^2/\hbar}\eta(g) du \,dY )$  , where $ u \in SU(2), g \in SL(2,C)$ 
This provides the following  $\tilde Y$ map of the functions of SU(2) to the square integrable functions of SL(2,C) with the above measure:
\begin{equation}
\label{tildey}
\tilde Y:  \phi(u) = \sum\limits_{j=0} \sum\limits_{q,q' = -j}^{j} C_{jqq'} D^j_{qq'}(u) \rightarrow \psi(g) = \sum\limits_{j=0} \sum\limits_{q,q' = -j}^{j} \frac{1}{A_j} C_{jqq'} D^{(j_-, j_+)}_{qq'}(g)
\end{equation}
As it was shown in \cite{Perlov2013}  the simplicity constraints provide the following solution for the spins and Barbero-Immirzi parameter: 
\begin{equation}
\label{gamma2}
\gamma = \frac{-in}{(|n|+2p)}, \; n \in Z, p = 0, 1, \mbox{...}
\end{equation}
\begin{equation}
 j_-=p/2, \; \; j_+= j+ p/2
\end{equation}
, $j_-$ and $j_+$ - are $SL(2,C)$ spinor representation parameters,  $j$ is an $SU(2)$ spin. 
After substituting it and $A_j$  from (\ref{holomorphicoeff}) into (\ref{tildey}) for the $\tilde Y$ map we obtain:
\begin{equation}
\label{tildey1}
\tilde Y:  \phi(u) = \sum\limits_{j=\frac{|m|}{2}} \sum\limits_{q,q' = -j}^{j} C_{jqq'} D^j_{qq'}(u) \rightarrow \psi(g) = \sum\limits_{j=\frac{|m|}{2}} \sum\limits_{q,q' = -j}^{j} \frac{1} {(\hbar\pi)}^{-3/4}e^{\frac{-\hbar{(2j + 1)}^2}{8}} C_{jqq'} D^{(p/2, j + p/2)}_{qq'}({g})
\end{equation}
Since the spinor representation is part of the non-unitary principal series representation with the parameters $(n, \rho), \; \; n \in Z, \rho \in C$, by expressing these parameters via the spins $(j_-, j_+)$ in the following way:
\begin{equation}
\label{nro}
n = (2j_+ - 2j_-),   \quad
i\rho = ( 2j_+ + 2j_-)
\end{equation}
or
\begin{equation}
(n, \rho = -i (|n| + 2p) )
\end{equation}
we can rewrite ($\ref{tildey1}$) and use the principal series matrix coefficients:
\begin{equation}
\label{tildey2}
\tilde{Y}:\phi(u) = \sum\limits_{j=\frac{|m|}{2}} \sum\limits_{q,q' = -j}^{j} C_{jqq'} D^j_{qq'}(u) \rightarrow \psi(g) = \sum\limits_{j=\frac{|m|}{2}} \sum\limits_{q,q' = -j}^{j}  {(\hbar\pi)}^{-3/4}e^{\frac{-\hbar{(2j + 1)}^2}{8}} C_{jqq'} D^{(\pm 2j, -2i( j + p)) }_{jqjq'}(\bar{g})
\end{equation}
where $j = |n|/2$ - $SU(2)$ spin.\\
We would like to emphasize that $A_j$ depends on j. 
According to the Holomorphic Huebschmann-Peter-Weyl theorem, the above sum is convergent and the function $\psi(g)$ is a square integrable function in $L^2(g, e^{-|Y|^2/\hbar} \eta(g) du \,dY ), g \in SL(2,C), u \in SU(2), Y \in su(2)$.\\[2ex]
Even though it first seems that the $Y$-map and  the $\tilde{Y}$-map both use the 3-dimensional projected inner product, in reality it is not the case. While in the $Y$-map definition one first selects the $SU(2)$ subgroup of $SL(2,C)$  to define the projected inner product on it, in $\tilde{Y}$-map one defines the inner product directly in 4 dimensional Lorentz space without breaking Lorentz covariance. Then it follows from the Holomorphic  Huebschmann-Peter-Weyl theorem  \cite{Huebschmann} that the inner product can be also calculated with the help of projections on $SU(2)$ subspaces and more than that, the projected inner product will be the same no matter what $SU(2)$ subspace we select. Therefore while the $Y$-map breaks the Lorentz covariance, the $\tilde{Y}$-map  preserves it. The real reason for this difference between the $Y$-map and  the $\tilde{Y}$-map lies in the fact that the $Y$-map image consists of non-square integrable functions on $SL(2,C)$ $\cite{Perlov2015-2}$, therefore it is impossible to  define the inner product directly, while the  $\tilde{Y}$ image consists of the square integrable functions  \cite{Huebschmann}. 

\section{Discussion}
\label{sec:Discussion}
In this paper we have introduced an alternative to the $Y$-map of a new LQG covariant formulation $\cite{RovelliBook2}$. We called a new map -  $\tilde{Y}$. We have shown that $\tilde{Y}$-map is well defined and convergent in the space of square integrable functions of $SL(2,C)$  with the measure $e^{-|Y|^2/\hbar}\eta(g) du \,dY$, as a consequence of  the Holomorphic Huebschmann-Peter-Weyl theorem \cite{Huebschmann}. The main difference between the $Y$-map and  $\tilde{Y}$-map is in Lorenz covariance and unitarity. The $Y$-map is unitary but breaks the Lorentz covariance, while the $\tilde{Y}$ map is non-unitary, but preserves the Lorentz covariance.  The unitary evolution in quantum gravity does not follow from the Wheeler-DeWitt dynamics equation $\hat{H}\Psi = 0$ as it follows from the Schrödinger equation in the classical quantum gravity, requiring that the inner product of the Hilbert space does not depend on time.  If one does not use the $SU(2)$ ADM-like slicing  by the 3-dimensional hypersurfaces then time is not distinguished from the space coordinates and it does not make any sense to consider evolution with respect to it. Therefore the choice of the non-unitary representation in quantum gravity is a valid choice.

\end{document}